\documentclass[%
 reprint,
superscriptaddress,
 amsmath,amssymb,
 aps,
 pra,
]{revtex4-1}

\usepackage{graphicx}% Include figure files
\usepackage{dcolumn}% Align table columns on decimal point
\usepackage{bm}% bold math
\usepackage{hyperref}% add hypertext capabilities
\usepackage{physics}
\usepackage{xcolor}

\begin{document}

\preprint{APS/123-QED}

\title{Hard-Wall Confinement of a Fractional Quantum Hall Liquid}% Force line breaks with \\

\author{E. Macaluso}
\affiliation{%
INO-CNR BEC Center and Dipartimento di Fisica, Universit$\grave{a}$ di Trento, 38123 Povo, Italy%\\
}%
\author{I. Carusotto}%
\affiliation{%
INO-CNR BEC Center and Dipartimento di Fisica, Universit$\grave{a}$ di Trento, 38123 Povo, Italy%\\
}%


%\affiliation{
%\affiliation{
 %Third institution, the second for Charlie Author
%\affiliation{%


\date{\today}% It is always \today, today,

\begin{abstract}
We make use of numerical exact diagonalization calculations to explore the physics of $\nu = 1/2$ bosonic fractional quantum Hall (FQH) droplets in the presence of experimentally realistic cylindrically symmetric hard-wall potentials. This kind of confinement is found to produce very different many-body spectra compared to a harmonic trap or the so-called extremely steep limit. For a relatively weak confinement, the degeneracies are lifted and the low-lying excited states organize themselves in energy branches that can be explained in terms of their Jack polynomial representation. For a strong confinement, a strong spatial deformation of the droplet is found, with an unexpected depletion of its central density.
\end{abstract}

\pacs{Valid PACS appear here}% PACS, the Physics and Astronomy
\maketitle

%\tableofcontents

\section{\label{sec:intro}INTRODUCTION}

Fractional quantum Hall (FQH) states have been first observed in two-dimensional electron gases in the presence of strong magnetic fields~\cite{Tsui}. Since then, they have been one of the most active branches of quantum condensed-matter physics, with a rich variety of intriguing phenomena and a close connection to the topological properties of the underlying many-body states~\cite{TongNotes,Goerbig}. Very recently, the interest for these systems has been renewed by long-term perspectives in view of quantum information processing applications~\cite{SarmaRMP}.

In parallel to these advances in electronic systems, fractional quantum Hall physics is receiving a strong attention also from the communities of researchers working on quantum gases of ultra-cold atoms~\cite{SSLP} and in quantum fluids of light~\cite{ICCC_RMP}. Even though they are electrically neutral particles, both atoms and photons can display magnetic effects when subject to the so-called synthetic or artificial magnetic fields. 

The first proposal in this direction has been to make a trapped atomic cloud to rotate at a fast angular speed and exploit the mathematical analogy between the Coriolis force and the Lorentz force on charged particles~\cite{Cooper}. Later on, researchers have focussed on dressing the atoms with suitably designed optical and magnetic fields so to associate a Berry phase to their motion~\cite{review_on_synthetic_atoms_Juzeliunas, goldman_juzeliunas}. 
In the last years, this has led to the observation of some among the most popular models of topological condensed-matter physics, such as the Hofstadter-Harper model~\cite{HofstadterColdAtoms,HofstadterKetterle}, the Haldane model~\cite{HaldaneColdAtoms}, as well as the so-called spin Hall effect ~\cite{SpinHallEffectAtoms} and the nucleation of quantized vortices by a synthetic magnetic field~\cite{Lin:Nature2009}. 

Also in the optical context, topologically protected edge states related to the integer quantum Hall ones have been observed in suitably designed magneto-optical photonic crystals~\cite{Wang:Nature2009}, in optical resonator lattices~\cite{hafezi} as well as in arrays of waveguides~\cite{rechtsman}, while non-planar macroscopic ring cavities have been demonstrated to support Landau levels for photons~\cite{simon}. 

In both atomic and optical systems, the present experimental challenge is to push the study of systems experiencing artificial magnetic fields into a regime of strongly interacting particles where strongly correlated states are expected to appear, in primis fractional quantum Hall states~\cite{TongNotes,Goerbig}. While too high values of the system temperature are one of the main difficulties encountered by atomic realizations, photonic systems are facing the challenges~\cite{IC_circumnavigating,ICCC_RMP} of generating sufficiently large photon-photon interactions mediated by the optical nonlinearity of the underlying medium and dealing with the intrinsically driven-dissipative nature of the photon gas. A promising solution to the former issue based on coherently dressed atoms in a Rydberg-EIT configuration has been investigated in~\cite{jia2017rydberg}. Different pumping schemes to generate quantum Hall states of light have been investigated in~\cite{Umucalilar:2012PRL,Kapit:2014PRX,OUIC_IncoherentFQH}.

With respect to electronic systems, atomic and photonic systems are expected to offer a much wider flexibility and a more precise control on the external potential confining the FQH droplet and, consequently, on the properties of its edge. Early theoretical works on these systems have focussed on harmonic confinements~\cite{PhysRevB.60.R16279, PhysRevLett.84.6, PhysRevLett.87.120405, PhysRevLett.91.030402, PhysRevB.71.121303} for which, however, the smoothness of the confining potential hinders a clear distinction between bulk and edge. Only recently researchers, motivated by the realization of a flat-bottomed traps for ultra-cold atoms \cite{Hadzibabic} and by the flexibility in designing optical cavities~\cite{simon} and arrays of them~\cite{roushan2016chiral}, have started investigating hard-wall (HW) potentials and the peculiar many-body spectral properties they induce in the excitation modes of the FQH droplet. Along these lines, a so-called extremely steep limit has been considered in \cite{FernSimon}, characterized by a marked hierarcy of the confinement potential experienced by the sequence of lowest-Landau-level single-particle orbitals. In particular, it was shown that the eigenstates of a $\nu = 1/r$ FQH droplet correspond under such a idealized confinement to certain Jack polynomials, from which one can extract an analytic expression for their energies.

In the present work, we provide a general study of the effect of a general and experimentally realistic HW potential on $\nu=1/2$ bosonic FQH droplets. Using a numerical approach based on exact diagonalization, we characterize the ground state of the confined system as well as its low-lying excited states depending on the confinement potential. 
In a weak confinement regime, a classification of excited states based on their representation in terms of Jack polynomials is proposed. The energy ordering of the branches and sub-branches of this spectrum and the relative energies of states within the same sub-branch are discussed and physically explained. This analysis confirms the presence of signicant deviations from the standard chiral Luttinger liquid theory of edge excitations~\cite{Xiao-GangWen} as first predicted in~\cite{FernSimon} for an idealized extremely steep limit, but also highlights crucial qualitative differences from this latter work. In the strong confinement regime, a peculiar deformation of the cloud with a marked density depletions at its center is pointed out and physically motivated.

The structure of the article is the following. In Sec.\ref{sec:model} we introduce the system Hamiltonian and we review the basics of Laughlin states and of their low-lying excitations in the unconfined and harmonic confinement cases. In Sec.\ref{sec:Jacks} we briefly review the main features of Jack polynomials and of their application to fractional quantum Hall physics. The main new results of this work are reported in Secs.\ref{sec:soft} and \ref{sec:strong} for the weak and the strong confinement cases, respectively. Conclusions are finally drawn in Sec.\ref{sec:conclusions}.

\section{\label{sec:model}Physical system and theoretical model}

\subsection{The model Hamiltonian}

We consider a 2D system described by the Hamiltonian $H = H_{0} + H_{int} + H_{conf}$, where
\begin{equation}
H_{0} = \sum^{\mathcal{N}}_{i=1} \frac{(\mathbf{p}_{i} + \mathbf{A})^{2}}{2 M}
\end{equation}
is the kinetic energy of particles experiencing an effective orthogonal magnetic field $\mathbf{B} = \mathbf{\nabla} \times \mathbf{A} = B \, \hat{e}_{z}$,
\begin{equation}
H_{int} = \sum_{i<j} g_{int} \, \delta^{(2)}(\mathbf{r}_{i} - \mathbf{r}_{j}),
\end{equation}
describes contact interactions between the particles of strength $g_{int}$ and finally
\begin{equation}
H_{conf} =  \sum^{\mathcal{N}}_{i=1}  V_{ext} \, \theta (|\mathbf{r}_{i}| - R_{ext}) 
\label{eq:confinement1}
\end{equation}
represents a radially symmetric hard-wall confining potential, which confines the particles in a disk-shaped region. Here, the potential height $V_{ext}$ and the disk radius $R_{ext}$ can set at will, while $\theta(x)$ denotes the Heaviside step function. 

For the sake of convenience, we choose the so-called symmetric gauge for the vector potential - i.e. $\mathbf{A} = B \, (-y/2,x/2,0)$ - which makes cylindrical rotational symmetry manifest and guarantees that all many-body eigenstates have a well-defined angular momentum.
For sufficiently large magnetic fields $B$, we can restrict ourselves to the lowest Landau level (LLL) and use the ladder operators $a^{\dagger}_{m}$ and $a_{m}$, which respectively create and annihilate particles in the LLL state of angular momentum $m$ and real-space wave function
\begin{equation}
\varphi_{m} (r, \phi) = \frac{1}{l_{B}\,\sqrt{2 \pi m!}} e^{i m \phi} \bigg( \frac{r}{\sqrt{2}l_{B}} \bigg)^{m} e^{- \frac{r^{2}}{4 l^{2}_{B}}}
\label{eq:LLL_wf}
\end{equation}
with magnetic length $l_{B}=\sqrt{\hbar c / B}$. 

This approximation is valid if we restrict to the low-energy physics of the system and we assume that the cyclotron energy $\hbar\omega_C=\hbar B/m$ is far larger than all other energy scales set by the potential energy $V_{ext}$ and the characteristic interaction energy $V_{0} \equiv g_{int}/ 2 l^{2}_{B}$. Within this approximation the Hamiltonian terms can be written in second-quantization terms as:
\begin{equation}
H_{0} = \varepsilon_{0} \sum_{m} a^{\dagger}_{m} a_{m}
\end{equation}
\begin{equation}
H_{int} = \frac{g_{int}}{2 \pi l^{2}_{B}} \sum_{\alpha \beta \gamma \rho} \frac{\Gamma(\alpha + \beta +1)}{\sqrt{\alpha ! \beta ! \gamma ! \rho !}} \frac{ \delta_{(\alpha + \beta , \gamma + \rho)}}{2^{(\alpha + \beta + 2)}} a^{\dagger}_{\alpha} a^{\dagger}_{\beta} a_{\gamma} a_{\rho}
\end{equation}
\begin{eqnarray}
H_{conf} &=&  \sum_{m} \mathcal{U}_{m} \, a^{\dagger}_{m} a_{m} \nonumber \\ &=& \sum_{m} \frac{V_{ext}}{m!}  \gamma_{\uparrow} \bigg( m +1, \frac{R^{2}_{ext}}{2l^{2}_{B}} \bigg) a^{\dagger}_{m} a_{m}
\label{eq:confinement2}
\end{eqnarray}
where $\varepsilon_{0} = \hbar \omega_{C} / 2$ is the kinetic energy of the (massively degenerate) LLL, $\Gamma (t)$ denotes the Euler gamma function
\begin{equation}
\Gamma (t) \equiv \int^{\infty}_{0} x^{t-1} e^{-x} \mathrm{d} x ,
\end{equation}
whereas $\gamma_{\uparrow}(t,R)$ is the so-called upper incomplete gamma function
\begin{equation}
\gamma_{\uparrow} (t,R) \equiv \int^{\infty}_{R} x^{t-1} e^{-x} \mathrm{d} x .
\end{equation}
Note that since we are neglecting excitation to higher Landau levels, all $\mathcal{N}$-particle states have the same kinetic energy $\varepsilon_{0} \, \mathcal{N}$, which effectively drops out of the problem. As a consequence we neglect the kinetic term and we focus on the Hamiltonian $\tilde{H} \equiv H - H_{0} = H_{int} + H_{conf}$.

%------------------------------------------------------------------------------------------------
\subsection{The confinement potential}

\begin{figure}[t]
\includegraphics[width=\columnwidth]{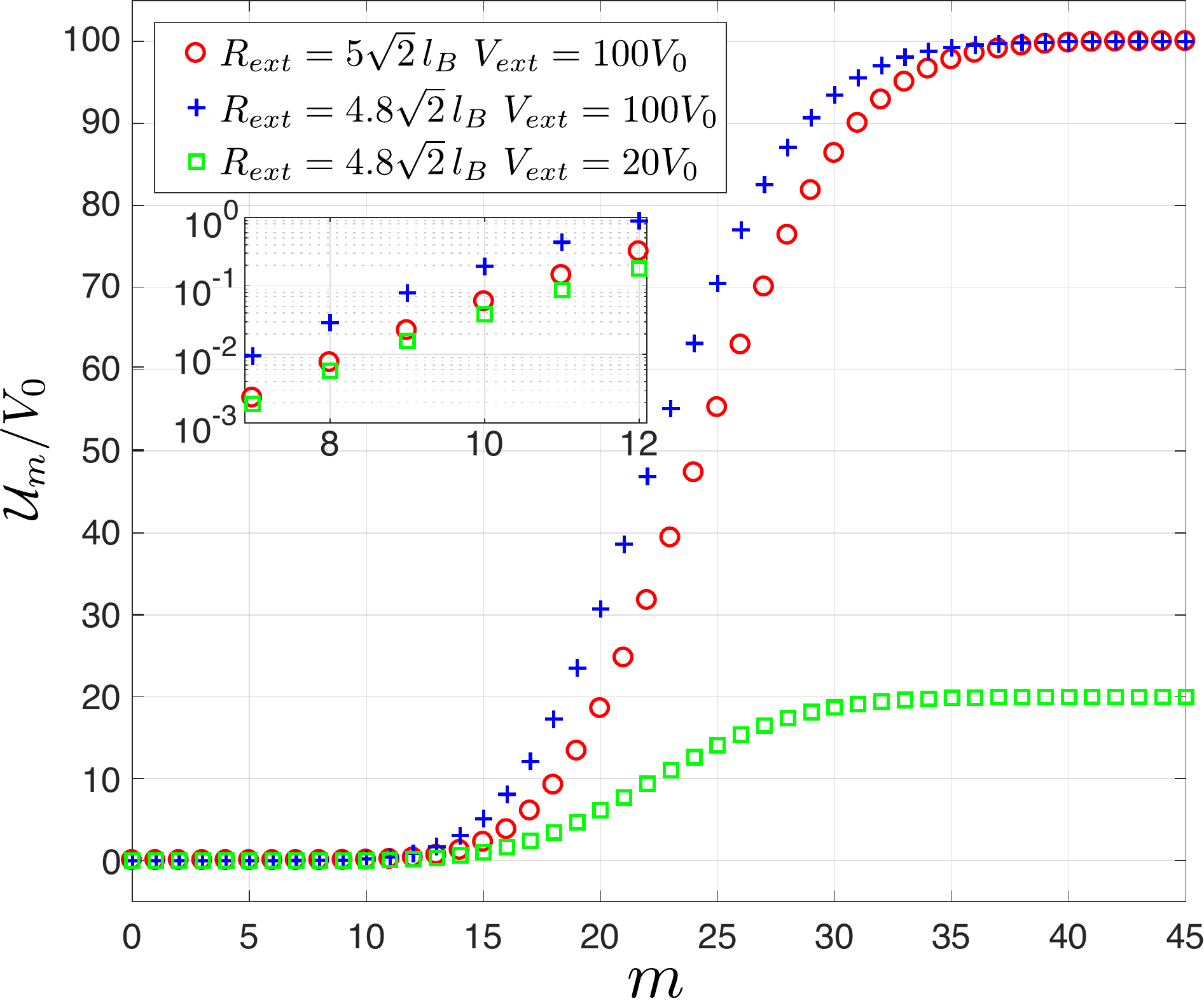}
\caption{Angular momentum-dependence of the confinement potentials $\mathcal{U}_m$.  The inset shows a magnified log-scale view on the region of $m$ values corresponding to the highest occupied single-particle orbitals in a $\mathcal{N} = 6$ Laughlin state. The potential parameters considered here are those typically used for the confinement of $\mathcal{N} = 6$ particle systems, for which the quite slow increase of the $\mathcal{U}_{m}$'s does not fulfill the extremely steep condition.}
\label{fig:potential_expansion}
\end{figure}

Within the LLL approximation, the confinement potential is summarized by its value $\mathcal{U}_m$ on each single-particle state of angular momentum $m$. This quantity is plotted in Fig. \ref{fig:potential_expansion} for a few choices of $V_{ext}$ and $R_{ext}$. In the recent work \cite{FernSimon}, this dependence was assumed to fulfill the condition $\mathcal{U}_{m-1} \ll \mathcal{U}_{m} \ll \mathcal{U}_{m+1} $, but in practice this condition does not appear to be so simple to satisfy with a realistic potential.

For sufficiently large $R_{ext}$, we can in fact approximate
\begin{equation}
 \mathcal{U}_m\simeq \frac{V_{ext}}{m!}\,\left(\frac{R_{ext}^2}{2l_B^2}\right)^m\,\exp\left(-\frac{R_{ext}^2}{2l_B^2}\right),
 \label{eq:Um}
\end{equation}
so that having a large ratio
\begin{equation}
 \frac{\mathcal{U}_m}{\mathcal{U}_{m-1}}\simeq \frac{R_{ext}^2}{2ml_B^2}
 \label{Um/Um-1}\end{equation}
requires a wide disk radius $R_{ext}\gg \sqrt{2m} l_B$, much larger than the FQH droplet we intend to prepare. On the other hand, for $\sqrt{2m} l_B \gtrsim R_{ext}$, the potential $\mathcal{U}_m$ smoothly approaches its limiting value $V_{ext}$. 

Given the exponential factor in \eqref{eq:Um}, simultaneously having an overall appreciable potential $\mathcal{U}_m$ and a very steep $m$-dependence (\ref{Um/Um-1}) requires a very large potential strength $V_{ext}$: physically, this is due to the fact that the HW potential only acts on the far tail of the LLL wave function. However, having a remote, but strong HW potential makes the system very sensitive to weak variations of the HW parameters. For instance, the relative variation of the confinement potential $\Delta \mathcal{U}_m/\mathcal{U}_m$ for a small variation $\Delta R_{ext}$ can be estimated to be a large number
\begin{equation}
\left|\frac{\Delta R_{ext}}{R_{ext}}\left[2m-\frac{R_{ext}^2}{2l_B^2}\right]\right|\simeq \frac{R_{ext}}{l_B}\,\frac{\Delta R_{ext}}{l_B}. 
\end{equation}
These arguments on the difficulty of fulfilling the condition assumed in \cite{FernSimon} are a further motivation for our numerical calculations including a realistic form of the confinement potential.

The numerical calculations reported in this work are based on a direct exact diagonalization (ED) of the second-quantized Hamiltonian on a truncated Fock space spanned by $\ket{n_{0}, n_{1}, n_{2}, \dots}$ number states with $n_{m} = 0, 1, \dots , \mathcal{N}$ particles in the $m$-th LLL orbital. To be precise, before diagonalizing the Hamiltonian matrix we set to zero all entries associated with basis elements having total number of particles lower than $\mathcal{N}$ and/or angular momentum quantum number different from those of interest.

Furthermore, we can take advantage of the description in terms of Jack polynomials \cite{DUMITRIU2007587, PhysRevLett.100.246802, PhysRevLett.103.206801, EdgeJacks} as an heuristic tool to conveniently choose the cutoff $m_{max}$ on the single particle LLL orbitals to be included in the calculation. Jack polynomials indeed allow to know the number of single-particle orbitals needed for the description of the Laughlin state and its edge and quasi-hole excitations in the absence of confinement. Since the external confinement has the effect of shrinking the cloud, we do not expect that the description of the eigenstates of the total Hamiltonian including confinement will require any additional orbital. As a further check, the accuracy of the numerical results has been ensured by verifying their independence on the cut-off.

%------------------------------------------------------------------------------------------------
\subsection{Laughlin states in the absence of confinement}

In absence of any confinement - i.e. $V_{ext} = 0$ and/or $R_{ext} \rightarrow \infty$ -, the eigenstates of a system of contact-interacting bosonic $2$D particles experiencing an effective orthogonal magnetic field are well-known from the theory. In particular, such unconfined system is characterized by a widely degenerate ground state, containing the $\nu = 1/2$ Laughlin state, as well as its quasi-hole (QH) and edge excitations (EEs). All such states are characterized by a vanishing interaction energy and are separated from excited states (including quasi-particle (QP) excitations) by a bulk excitation gap proportional to the interaction energy $V_0$ -- the so-called Laughlin gap.

In more detail, the $\nu = 1/2$ Laughlin state has the lowest angular momentum value $L_{L} = \mathcal{N} (\mathcal{N}-1)$ among all the states forming such a degenerate ground state and is described by the celebrated wave function:
\begin{equation}
\psi_{L}(\{ z_{i}\}) \propto \prod_{i<j} (z_{i} - z_{j})^{2} \, e^{- \sum_{i} |z_{i}|^{2} / 4 l^{2}_{B} } ,
\label{eq:Laughlin_wf}
\end{equation}
in which $z_{k} = x_{k} - i \, y_{k}$ denotes the position of the $k$-th particle in the complex plane.
In their general form, the wave functions of states in the ground state manifold can be written as
\begin{equation}
\psi_{edge} (\{z_{i} \}) \propto S (\{ z_{i} \}) \, \psi_{L} (\{z_{i} \}) ,
\label{eq:edge_wf}
\end{equation}
where $\psi_{L} (\{ z_{i} \})$ denotes the Laughlin wave function \eqref{eq:Laughlin_wf} and $S (\{ z_{i} \})$ is a generic homogeneous symmetric polynomial in the particle coordinates whose degree gives the additional angular momentum $\Delta L$. The degeneracy of such states is given by the number of partitions of the integer $\Delta L$. 

For low additional angular momentum $\Delta L\ll \mathcal{N}$, the edge excitations (EE) can be understood as area-preserving shape deformations of the FQH droplet~\cite{Cazalilla}. Another remarkable class of states among those in \eqref{eq:edge_wf} are quasi-hole (QH) excitations, characterized by fractional charge and anyonic statistics~\cite{TongNotes} and corresponding to density depletions of a half of a particle within a Laughlin state. In general a $\nu = 1/2$ Laughlin state presenting $n$ QH excitations at positions $\xi_{1}, \dots , \xi_{n}$ is described by the wave function:
\begin{equation}
\psi_{n-qh} (\{ z_{i}\} ,\{\xi_{i} \}) \propto \bigg( \prod^{\mathcal{N}}_{i=1} \prod^{n}_{j=1} (z_{i} - \xi_{j}) \bigg) \psi_{L}(\{ z_{i} \}) .
\end{equation} 
In the following, in order to preserve radial symmetry, we will restrict ourselves to states presenting QHs in the origin $\xi_{i} = 0$ for all $i = 1, \dots, n$, so that
\begin{equation}
\psi_{n-qh,0} (\{z_{i} \}, \{\xi_{i} = 0 \}) \propto \bigg( \prod_{i} z^{n}_{i} \bigg) \psi_{L} (\{ z_{i} \}).
\label{eq:n-QH_wf}
\end{equation}

\begin{figure}
\includegraphics[width=0.515\textwidth]{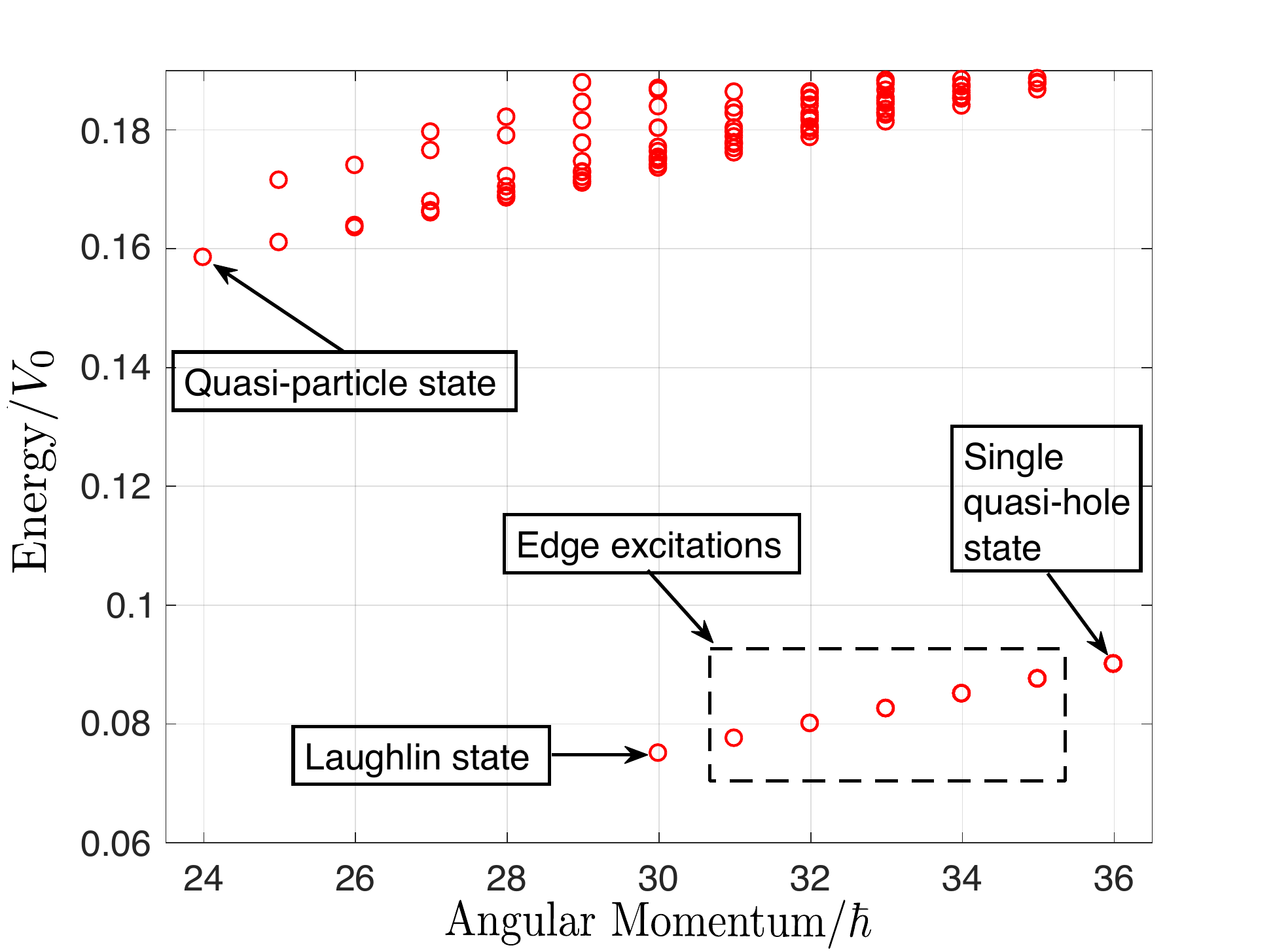}
\caption{Spectrum of many-body states of a $\mathcal{N} = 6$ particle system experiencing a harmonic confinement of frequency $\hbar \upsilon = 2.5 \times 10^{-3} \, V_{0}$. While the ground state is the Laughlin state, the energies of the low-lying excited states is proportional to the total angular momentum: all EEs with the same angular momentum are thus degenerate and the dynamics shows a single characteristic frequency $v$. The width of the bulk Laughlin gap to the lowest quasi-particle states is controlled by the interaction energy $V_0$.}
\label{fig:harmonic-spectrum}
\end{figure}

In the absence of confinement, all these QHs and EEs are degenerate with the Laughlin ground state. On the other hand, it is known that the introduction of an external confining potential allows to at least partially remove this degeneracy, in a way which depends on both the potential shape and on the confinement parameters. For instance, a harmonic potential [see Fig. \ref{fig:harmonic-spectrum}] introduces an energy shift $\Delta E = \upsilon L_{z}$ proportional to the angular momentum and to the harmonic confinement frequency. As a consequence, this confinement is able to remove the degeneracy between states with different angular momentum and to provide a non degenerate ground state in the Laughlin form \eqref{eq:Laughlin_wf}. However, excited states with the same angular momentum remain degenerate.

As we shall see in the following of this work, the hard-wall confinement \eqref{eq:confinement1} is instead able to completely remove the degeneracy of the lowest energy states of a $\nu=1/2$ FQH liquid and to induce a rich structure in the energy vs. angular-momentum diagram of many-body states. To capture the physics underlying this organization, Jack polynomials will be an essential tool. 

%------------------------------------------------------------------------------------------------
%------------------------------------------------------------------------------------------------
\section{\label{sec:Jacks}Basics of Jack polynomials}

In the previous section, we have briefly mentioned Jack polynomials - or simply \emph{Jacks} - as a useful tool to properly choose the cut-off on the single-particle orbitals to be included in the numerical calculation wave functions. As we shall see in the following, their power goes far beyond this as they allow to easily build useful trial wave functions to describe FQH states. In this section we shall review those main features that will be used in Sec.\ref{sec:soft} for the interpretation of the energy spectra of $\nu =1/2$ FQH droplets.

In general, Jacks $J^{\alpha}_{\lambda}$ are homogeneous symmetric polynomials identified by a rational parameter $\alpha$ - called \emph{Jack parameter} - and by a root partition $\lambda$. Here, by partition we mean a non-growing sequence of positive integers, $\lambda=[\lambda_1,\lambda_2,\ldots]$, so that the sum of the integers in the sequence corresponds to the number that gets partitioned. The degree of the Jack is given by this number, indicated by $|\lambda|$. Furthermore, 
Jacks have been found to exhibit clustering properties \cite{PhysRevB.77.184502} and also to correspond to some of the polynomial solution of the so-called \emph{Laplace-Beltrami operator} \cite{STANLEY198976}:
\begin{equation}
H^{\alpha}_{LB} = \sum_{i} ( z_{i} \partial_{i} )^{2} + \frac{1}{\alpha} \sum_{i<j} \frac{z_{i} + z_{j}}{z_{i} - z_{j}} ( z_{i} \partial_{i} - z_{j} \partial_{j} ) ,
\label{eq:Laplace-Beltrami-op}
\end{equation}
where $\partial_{i} \equiv \partial / \partial z_{i}$.

Partitions of length $N$ can be biunivocally associated to symmetrized monomials in $\mathcal{N}\geq N$ variables in the following way: the partition $[\lambda_1,\lambda_2,\lambda_3,\ldots,\lambda_N]$ corresponds to the monomial $\textrm{Sym}(z_1^{\lambda_1} z_2^{\lambda_2} z_3^{\lambda_3}\ldots z_N^{\lambda_N})$ where all other variables $z_{N+1}\ldots z_{\mathcal{N}}$ appear with degree 0. For example, the partition $[4,2,2,1]$ corresponds to the symmetrized monomial $ \mathcal{M}_{[4,2,2,1]} \equiv \mathrm{Sym}(z^{4}_{1} z^{2}_{2} z^{2}_{3} z^{1}_{4})$. In this perspective, Jacks have a peculiar expansion in terms of symmetrized monomials $\mathcal{M}_{\mu}$'s:
\begin{equation}
J^{\alpha}_{\lambda} = \mathcal{M}_{\lambda} + \sum_{\mu \succ \lambda} c_{\lambda \mu} (\alpha) \mathcal{M}_{\mu} ,
\label{eq:Jack_expansion}
\end{equation}
where $\mu$ runs over all partitions that can be obtained from the root partition $\lambda$ through all possible sequences of \emph{squeezing operations} \cite{PhysRevB.77.184502}. Under such an operation, one starts from a \emph{parent} partition $[\dots,\lambda_{i},\dots,\lambda_{j}, \dots]$ to generate another one - called \emph{descendant} - of the form $[\dots,\lambda_{i} - \delta m,\dots,\lambda_{j} + \delta m, \dots]$ (with $\lambda_{i} - \delta m \geq \lambda_{j} + \delta m$). The corresponding coefficients $c_{\lambda \mu} (\alpha)$ can be computed by means of a recursive construction algorithm \cite{DUMITRIU2007587, PhysRevB.84.045127}.

Jack polynomials with negative $\alpha$ appear in the theory of the quantum Hall effect. In this context the admissible root configurations are given by some of the so-called ($k,r$) admissible root configurations. The ($k,r$) admissibility means that there can not be more than $k$ particles into $r$ consecutive orbitals. In particular the bosonic Read-Rezayi $k$ series of states has been proven  to be given by single Jacks of parameter $\alpha = - \frac{k+1}{r-1}$ and suitable root partition~\cite{PhysRevLett.100.246802}. Among these bosonic FQH states, we focus here on the $\nu = 1/2$ Laughlin state for which $(k=1, r=2)$ and the wave function \eqref{eq:Laughlin_wf} turns out to be given by the Jack polynomial with $\alpha = -2$ and root partition $\lambda=[2 \mathcal{N} - 2, 2 \mathcal{N} - 4, \dots, 2]$.

To better understand this connection, it is useful to establish a link between a partition and a distribution of particles among the different LLL orbitals. Neglecting the ubiquitous Gaussian factor and the normalization constant, each LLL single-particle wave function \eqref{eq:LLL_wf} of angular momentum $m$ corresponds to a monomial $z^{m}$. So, any bosonic many-particle state $\ket{n_{0}, n_{1}, n_{2}, \dots}$ obtained by occupying each LLL wave function of angular momentum $m$ with a well-defined number $n_m$ particles can be described as a symmetrized monomial $\mathcal{M}_{\lambda}$, where $\lambda \equiv [\lambda_{1}, \lambda_{2}, \dots]$ is a sequence of positive integers -- i.e. a partition -- which indicates the LLL wave functions occupied by the different particles in descending order. For example, states of $\mathcal{N} \geq 4$ particles having one particle in the $m=4$ orbital, two in the $m=2$ orbital, one in the $m=1$ orbital and the remaining in the $m=0$ orbital can be written as $\ket{\mathcal{N}-4,1,2,0,1} = \mathrm{Sym}(z^{4}_{1} z^{2}_{2} z^{2}_{3} z^{1}_{4}) \equiv \mathcal{M}_{[4,2,2,1]}$.

\subsection{Laughlin states and their excitations in terms of Jacks}

Based on this connection, all rotationally symmetric eigenstates of the Hamiltonian
\begin{equation}
\ket{\psi} = \sum_{i} c_{i} \ket{n_{0}, n_{1}, n_{2}, \dots}_{i},
\label{eq:linear_comb}
\end{equation}
correspond to homogeneous symmetric polynomials in the particle coordinates of degree equal to the well-defined total angular momentum 
\begin{equation}
L=\sum_{m} m \, \bar{n}_{m},
\label{eq:fixedL}
\end{equation} 
where $\bar{n}_{m} \equiv \expval{a^{\dagger}_{m} a_{m}}{\psi}$.
In addition, one can check that the Laughlin state \eqref{eq:Laughlin_wf} - neglecting the Gaussian term - satisfies \eqref{eq:Laplace-Beltrami-op} for $\alpha = -2$, which reflects the fact that \eqref{eq:Laughlin_wf} vanishes as $(z_{i} - z_{j})^{2}$ when the $i$-th and the $j$-th particles approach each other. In particular, one can expand $\prod_{i<j} (z_{i} - z_{j})^2$ and check that the obtained expansion is exactly the one in \eqref{eq:Jack_expansion} with root partition $\Omega = [2 \mathcal{N} - 2, 2 \mathcal{N} - 4, \dots, 2]$. This means that by knowing the Jack parameter $\alpha=-2$ and the root configuration $\mathcal{M}_{\Omega} \equiv \ket{1 \, 0 \, 1 \, 0 \dots 1 \, 0 \, 1}$ one can construct the full Laughlin state simply applying the squeezing operation and the recursive formula for the coefficients.

This formulation in terms of Jacks can be extended to both QH and EE excited states above the Laughlin state. 
Concerning EEs, studies have demonstrated that in general their wave functions \eqref{eq:edge_wf} can not be expressed as single Jacks, but are linear combinations of a finite set of them \cite{EdgeJacks}. In particular, edge states of angular momentum $L = L_{L} + \Delta L$, can be expressed as linear combination of those $\alpha = -2$ Jacks whose partitions are obtained by adding to the root partition $\Omega$ of the Laughlin state any partition $\eta$ of the additional angular momentum $|\eta| = \Delta L$ of maximum length equal to the number $\mathcal{N}$ of particles, namely $\lambda = \Omega + \eta$. Here, the sum of two partitions $\lambda=[\lambda_1\ldots\lambda_l]$ and $[\mu_1\ldots \mu_m]$ of lengths $l\geq m$ is defined as a partition of length $l$ whose elements are $\lambda+\mu=[\lambda_1+\mu_1\ldots\lambda_m+\mu_m,\lambda_{m+1}\ldots\lambda_l]$ (extension to the $l<m$ case is straightforward by requiring the sum to be a commutative operation).

Is then immediate to check that the Jack polynomials obtained in this way correspond to partitions satisfying $|\lambda| = |\Omega| + |\eta| = L_{L} + \Delta L = L$, so the corresponding states indeed have the required angular momentum. Furthermore, this choice for the possible \emph{edge partitions} (EPs) $\eta$'s is in agreement with the degeneracy predicted by the single branch chiral Luttinger theory of edge states \cite{Xiao-GangWen}. 

As an example, consider the $\Delta L = 2$ edge excitations of the $\mathcal{N} = 6$ particle $\nu = 1/2$ Laughlin state. In this case there are two possible EPs given by $\eta^{a} = [2]$ and $\eta^{b} = [1,1]$. Therefore the $\Delta L = 2$ edge states trial wave functions can be obtained as linear combinations of Jacks $J^{(-2)}_{\lambda^{a}}$ and $J^{(-2)}_{\lambda^{b}}$, where the admissible root partitions are $\lambda^{a} = \Omega + \eta^{a} = [10,8,6,4,2] + [2] = [12,8,6,4,2]$ and $\lambda^{b} = \Omega + \eta^{b} = [10,8,6,4,2] + [1,1] = [11,9,6,4,2]$. In terms of root configurations this means moving particles occupying the highest-$m$ occupied orbitals into orbitals with even higher single-particle angular momenta. In the concrete example above, $\lambda^{a}$ is obtained by moving one particle from the $m=10$ orbital to the $m=12$ orbital, while $\lambda^{b}$ is obtained by moving one particle from the $m=10$ orbital to the $m=11$ one and another from the $m=8$ orbital to the one with $m=9$.

The situation is simpler for the state displaying $n$ QH excitations located at $z=0$, whose wave function is given in \eqref{eq:n-QH_wf}. In contrast to the generic EEs considered above, this state can be expressed as a single $\alpha=-2$ Jack polynomial: the root configuration for the $n$-QH wave function \eqref{eq:n-QH_wf} is given by $\ket{0^{n} 1 \, 0 \, 1 \dots 1 \, 0 \, 1}$, which can be obtained starting from the Laughlin one by moving each particle from the $m$ orbital to the $m+n$ orbital. The root partition corresponding to such a configuration is therefore $\lambda^{n} \equiv \Omega + \kappa^{n}$, where $\Omega$ denotes again the root partition associated with the $\nu=1/2$ Laughlin state and the \emph{n-QH partition} $\kappa^{n}$ is the sequence obtained by repeating $\mathcal{N}$ times the same integer $n$, i.e. $\kappa^{n} \equiv [n, n, \dots, n] $.

\subsection{Edge Jacks}

The recent work \cite{EdgeJacks} has proven an interesting relation between Jacks with EPs $\eta$'s as their roots - which we will call \emph{edge Jacks} (EJs) - and the Jacks with root partitions of the form $\lambda = \Omega + \eta$ discussed in the previous subsection. In particular, they are related by
\begin{equation}
J^{\alpha}_{\Omega + \eta} = 
J^{\beta}_{\eta} J^{\alpha}_{\Omega} ,
\label{eq:Jack_relation}
\end{equation}
in which $\alpha = - \frac{2}{r-1}$, $\beta = \frac{2}{r +1}$ and $J^{\alpha}_{\Omega}$ denotes the Jack representing the $\nu = 1/ r$ Laughlin wave function. In our specific case of $\nu = 1/2$ Laughlin states, we have $\beta = 2/3$. 

Relation \eqref{eq:Jack_relation} suggests a description of generic edge states described by a homogeneous symmetric polynomial pre-factors $S(\{z_{i}\})$ in \eqref{eq:edge_wf} in terms of Jacks. The polynomial $S(\{z_{i}\})$ can in fact be expanded in the basis of Jacks with Jack parameter $\beta = 2/3$ and different partitions $\eta$'s. Each of these edge state wave functions $J^{\beta}_{\eta} \, \psi_{L} \{z_{i}\}$ can then be written as a single Jack using \eqref{eq:Jack_relation} and therefore they can be easily constructed using the expansion \eqref{eq:Jack_expansion},
 \begin{equation}
J^{-2}_{\Omega + \eta} = J^{2/3}_{\eta} \, J^{-2}_{\Omega} = J^{2/3}_{\eta} \, \psi_{L} (\{z_{i}\}).
\label{eq:edgeJacks}
\end{equation}
A drawback of this procedure is that wave functions of this form for different edge partitions $\eta$'s are not orthogonal for $\beta = 2/3$.

\begin{figure*}[htb]
\includegraphics[width=0.995\textwidth]{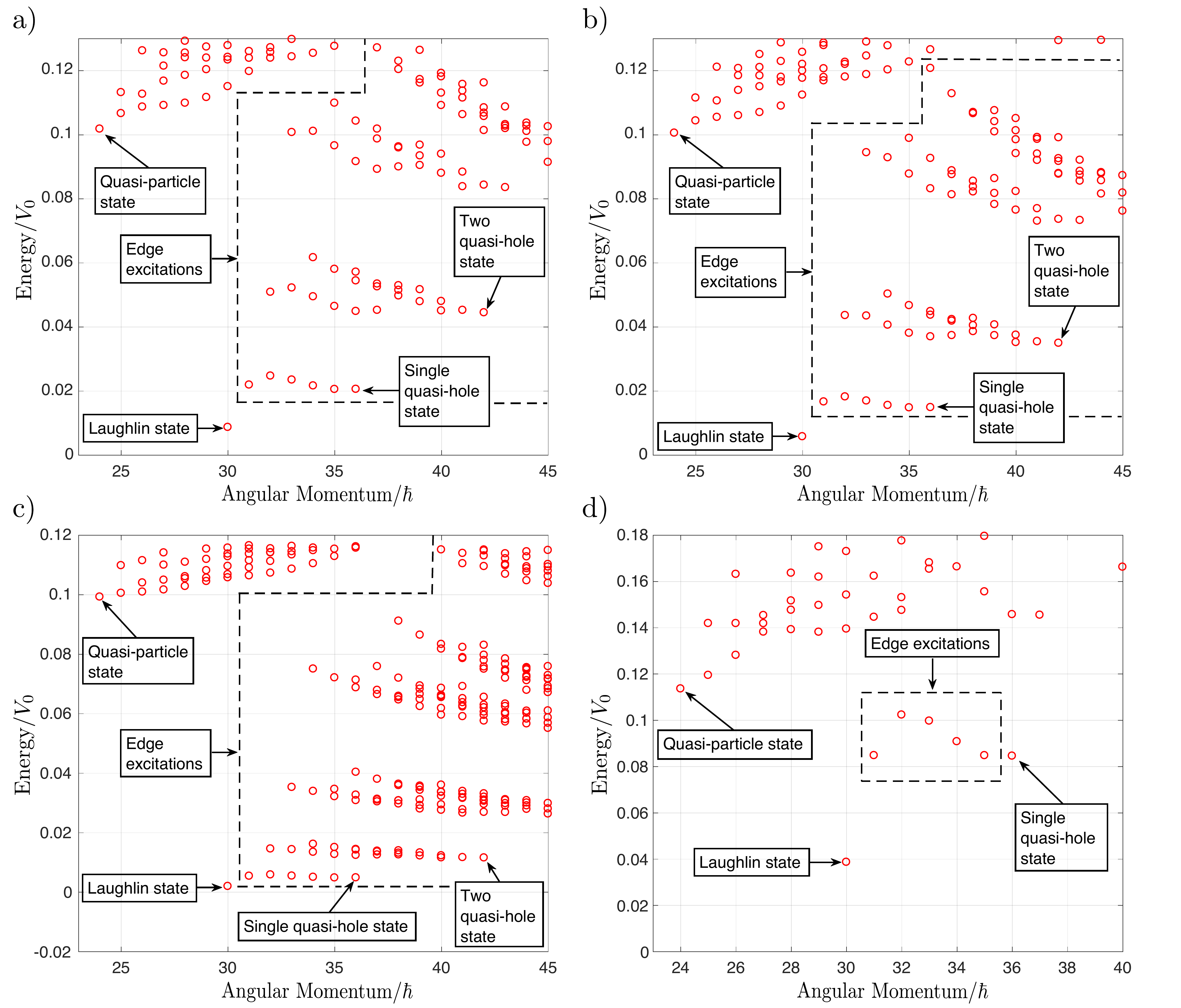}
\caption{Many-body spectra for $\mathcal{N} = 6$ particle system in the weak confinement regime experiencing different HW confining potentials. While the chiral Luttinger liquid theory well captures the number of energy branches and sub-branches for each value of $\Delta L$, the characteristic linear relation between energy and angular momentum breaks down for all choices of HW parameters.Panels a) and b) correspond to a HW confinement with $V_{ext} = 20 \, V_{0}$ and $R_{ext} = 4.95 \sqrt{2} \, l_{B}$ and with $V_{ext} = 100 \, V_{0}$ and $R_{ext} = 5.25 \sqrt{2} \, l_{B}$, respectively. In both cases the ground state is given by the Laughlin one and for a fixed value of $\Delta L$ different non-degenerate EEs can be resolved below the bulk excitation gap. In these spectra also the two different sub-branches of the second EE energy brach can be easily distinguished. Panel c) corresponds to an even weaker HW confinement with $V_{ext} = 30 \, V_{0}$ and $R_{ext} = 5.25 \sqrt{2} \, l_{B}$: in this case all low-lying excited states up to $L = 34$ have energies below the bulk excitation gap so that the first four energy branches are visible. Panel d) corresponds to a stronger HW confinement with $V_{ext} = 100 \, V_{0}$ and $R_{ext} = 4.95 \sqrt{2} \, l_{B}$ for which the ground state and first excited one are given by the Laughlin state and the single-QH state, respectively. In this case, for a given value of $\Delta L < \mathcal{N}$ there is just one non-degenerate branch of excited states below the bulk excitation gap.}
\label{fig:soft_spectra}
\end{figure*}

This potential difficulty can be overcome by using an alternative expansion of the $S(\{z_{i}\})$ pre-factors on the different basis of EJs of the form $J^{\beta=\nu}_{\eta}$, where $\nu = 1/r$ is the usual FQH filling factor and $\eta$ is one of the EPs discussed above~\cite{FernSimon}. Since such Jacks with $\beta=\nu$ are orthogonal in the thermodynamic limit,
\begin{equation}
\braket{J^{\nu}_{\eta}}{J^{\nu}_{\mu}} = j_{\mu}(\nu) \, \delta_{\eta , \mu}  \quad \mathrm{for} \quad \mathcal{N} \rightarrow \infty,
\end{equation}
under the Laughlin scalar product
\begin{equation}
\braket{\phi}{\chi} \equiv \int_{\mathbb{C}^{\mathcal{N}}} \mathrm{d} \{z_{i}\} [\phi(\{z_{i}\})]^{*}\chi(\{z_{i}\}) \, |\psi_{L} (\{z_{i}\})|^{2} ,
\end{equation}
the associated edge state wave functions turn out to be also orthogonal in the limit of $\mathcal{N} \rightarrow \infty$. While use of these $\beta=\nu$ EJs $J^{\nu}_{\eta}$ has the great advantage of leading to edge state wave functions that are orthogonal in the thermodynamic limit, its drawback is that these wave functions in general can not be written as single Jacks - as it instead happens for $\beta = \frac{2}{r +1}$ EJ's via eq.(\ref{eq:Jack_relation}). In the next section we will make use of these wave functions to interpret the result of our numerical ED calculations.

%------------------------------------------------------------------------------------------------
%------------------------------------------------------------------------------------------------
\section{\label{sec:soft}Weak confinement regime}

After reviewing the basic concepts of fractional quantum Hall physics and of Jack polynomials, in the next two sections we are going to present and discuss our ED numerical results for the low-lying part of the many-body spectrum for different values of the HW potential parameters $V_{ext}$ and $R_{ext}$. Different regimes can be distinguished depending on the value of these parameters, in particular we identify a \emph{weak confinement regime} (discussed in this section) and a \emph{strong confinement regime} (considered in the next Sec.\ref{sec:strong}).
In both cases we restrict to the $R_{ext} \gtrsim R_{cl} $ case, where $R_{cl} \simeq \sqrt{\mathcal{N} / \nu} \, \sqrt{2} l_{B}$ denotes the semiclassical radius of a $\mathcal{N}$-particle FQH droplet with filling factor $\nu$. While this assumption guarantees that the $\mathcal{U}_m$ components of the confinement potential are a growing function of $m$, it includes a wider set of potentials than the extremely steep HW limit of~\cite{FernSimon}.

The \emph{weak confinement regime} is characterized by a weak mixing of the Laughlin state and its EE and QH excited states with states above the Laughlin gap, e.g. quasi-particle excitations. In this regime, as one can see in Fig. \ref{fig:soft_spectra} the ground state of the system is non-degenerate and is very close to the $\nu = 1/2$ Laughlin state \eqref{eq:Laughlin_wf}.
On the other hand, the HW potential pushes the low-lying excited states to higher energies and removes the degeneracy between EEs with the same additional angular momentum $\Delta L$. In particular, for a given $\Delta L$, the number of EEs with energies lying below the bulk excitation gap depends on the potential parameters. Indeed more we increase the potential strength --or we reduce the potential radius-- more EEs end up having energies lying above the bulk Laughlin excitation gap. As a consequence, only for the lowest values of $\Delta L$ it is possible to resolve all EEs [see Fig. \ref{fig:soft_spectra} a)-c)]. 

As a first step of our study of the FQH liquid under a weak HW confinement, in Sec.\ref{subsec:classification}, we will present a classification of the many-body states into branches and sub-branches and we will highlight its relation to the EP of the corresponding Jack polynomials along the lines of~\cite{FernSimon}: the excellent overlap with the analytic trial wavefunction is remarkably associated to significant deviations from the chiral Luttinger liquid theory of the edge, as visible in both the degeneracy of states and in their ordering within a given sub-branch. More details on the ordering of the different sub-branches are given in Sec.\ref{subsec:sub-branches}: discrepancies from the extremely steep limit of~\cite{FernSimon} are pointed out and a unexpected energy-crossing between states sharing the same total angular momentum highlighted for varying trap parameters. In the following Sec.\ref{subsec:QH}, we shall discuss how the dispersion of states within a given subbranch can be widely controlled via the HW potential parameters. As a striking example, we shall present a regime in which the single-QH state is the first excited state.  Finally, in \ref{subsec:compressibility}, the compressibility of the FQH liquid and of its quasi-hole and quasi-particle excited states is characterized in terms of the dependence of the eigenstate energies on the confinement potential strength.

\begin{figure*}
\includegraphics[width=1.0\textwidth]{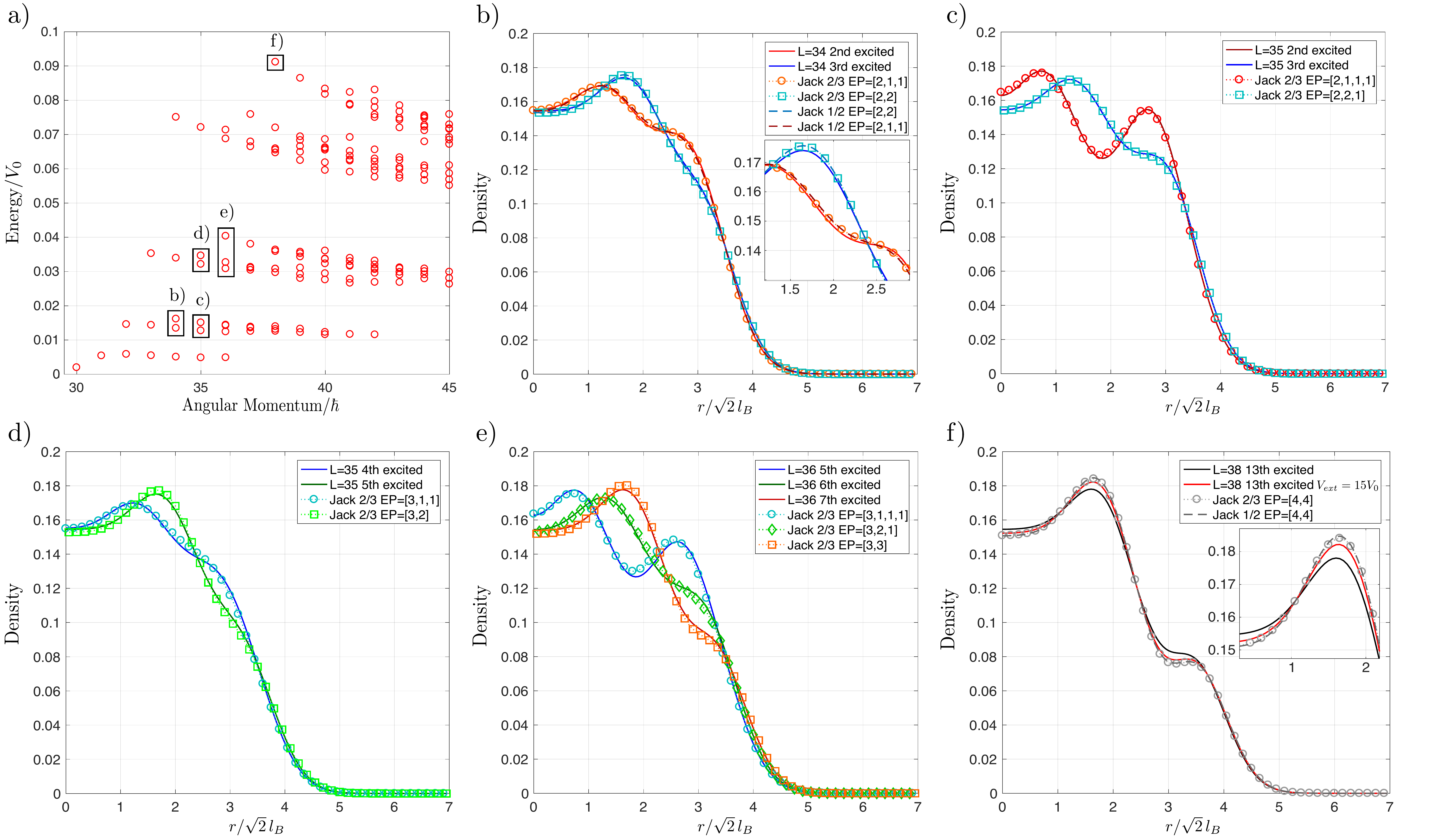}
\caption{a) Eigenstates of a $\mathcal{N} = 6$ particle system experiencing an HW potential of parameters $V_{ext} = 30 \, V_{0}$ and $R_{ext} = 5.25 \sqrt{2} \, l_{B}$ with energies lying below the bulk excitation gap and angular momenta $L \leq 45$. b)-f) Comparison between the density profiles of some of the eigenstates depicted in a) and the associated trial wave functions constructed starting from EJs of parameter $\nu = 1/2$ - dashed lines - and/or $\beta = 2/3$ - dotted lines with markers -. As we can see, the eigenstate description in terms of Jacks is so good that discrepancies in the density profiles are hard to see in most cases. In order to display how deviations of the numerical eigenstates from the Jacks trial wave functions can be further reduced by considering weaker confinements, panel f) shows also the density profile - red solid line - for a lower value of the HW strength, i.e. $V_{ext} = 15 V_{0}$. Insets in b) and f) show that on these scales differences between trial wave functions associated to the same EP but a different Jack parameter are not visible.}
\label{fig:soft_dens}
\end{figure*}

%------------------------------------------------------------------------------------------------
\subsection{\label{subsec:classification}Classification of states in terms of Jacks}

The global organization of the EEs is therefore easiest understood in Fig. \ref{fig:soft_spectra} c) where the confinement is weakest: looking at the full spectrum, instead of focusing on a specific value of $\Delta L$, we can see that EEs organize themselves in a sort of energy branches: the $k$-th branch starts from an angular momentum $L = L_{L} + k$ and ends with a state presenting $k$ QHs at the origin. In between, it splits into $k$ sub-branches.

Although such a structure of the spectrum could seem quite complicated, it can be completely explained in terms of Jack polynomials. In this and the next subsections, we are in fact going to see that in the weak confinement regime the EEs turn out to be well described by the wave functions
\begin{equation}
\phi_{\eta} (\{z_{i}\}) = J^{1/2}_{\eta} \, \psi_{L} (\{z_{i}\}),
\label{eq:Simon_wf}
\end{equation}
and the $k$-th energy branch contains all states whose wave functions can be obtained by multiplying $\psi_{L} (\{z_{i}\})$ by Jacks with partitions of the form
\begin{equation}
\eta = [ \eta_{1}, \eta_{2}, \eta_{3}, \dots] = [k, \eta_{2}, \eta_{3}, \dots].
\label{eq:k_edge_partition}
\end{equation}
As a first step, simple angular momentum arguments based on the observed extension of the branch support this statement. As a Jack of partition $\lambda = \Omega + \eta$ has angular momentum $L = L_{L} + |\eta|$, the lowest $\Delta L$ allowed by the form \eqref{eq:k_edge_partition} is the one associated with $\eta = [k]$, namely $\Delta L = k$. At the same time partitions are ordered sequences of positive integers and therefore they must satisfy
\begin{equation}
\eta_{1} \geq \eta_{2} \geq \dots \geq \eta_{\mathcal{N}}.
\label{eq:eta_i_relations}
\end{equation}
This implies that the maximum additional angular momentum that can be obtained by considering EPs with $\eta_{1} = k$ is $\Delta L = {\mathcal{N}}k$ and that it comes from the partition $[k, \dots, k]$ corresponding to the $k$-QH state.

Within this picture, the monotonically growing energy of the bands as a function of $k$ is easily understood as the main contribution to the confinement energy is given by the outermost particle, whose LLL wave functions is peaked on a larger ring. On the $k$-th branch, the outermost occupied orbital of the configuration is indeed the one of angular momentum $\Omega_{1} + k$, namely the one associated to the particle that was moved by $k$ orbitals in the outward direction from the outermost occupied orbital of the Laughlin root configuration.

Along the same lines, the order of the sub-branches in energy can be related to the value of the second element $\eta_{2}$ of the partition. In particular the $j$-th sub-branch of the $k$-th energy branch is composed of the states whose trial wave functions can be constructed from Jacks having partitions
\begin{equation}
\eta = [ \eta_{1}, \eta_{2}, \eta_{3}, \dots] = [k, j, \eta_{3}, \dots] ,
\end{equation}
with the exception of the $1$-st sub-branch in which there is also the Jack with partition $\eta = [k]$ having $\eta_{1} = k$ and no $\eta_{2}$.  This interpretation of the different sub-branches, together with the constraint \eqref{eq:eta_i_relations}, is further confirmed by the fact that the number of states in a given sub-branch depends only on the sub-branch index $j$ and on the number of particles $\mathcal{N}$ - which fixes the maximum number of partition elements $\eta_{i}$ -, but not on the branch index $k$.
Subtle features related to this ordering of levels will be highlighted in more detail in Sec.\ref{subsec:sub-branches} and compared to the conclusions of~\cite{FernSimon} for the extremely steep HW limit.

Our interpretation of the branches is numerically validated in the plots of Fig.\ref{fig:soft_dens}, which successfully compare the density profiles predicted by the trial wave functions in \eqref{eq:edgeJacks} to the one numerically predicted by ED. The deviations are a consequence of the mixing with quasi-particle excitations induced by the confinement as well as of the non-orthogonality of wave functions associated with different EPs of the same additional angular momentum $\Delta L$. Note that the calculation of the density profiles shown in this picture is made significantly simpler by the expression of the edge state wavefunction in terms of single Jacks via \eqref{eq:edgeJacks}.

In some panels, we have added the corresponding density profiles calculated using the trial wave function \eqref{eq:Simon_wf} (in others, the curves for the two kinds of Jacks are indistinguishable on the scale of the figure). The agreement with the numerics is once again excellent. As these trial wave functions are orthogonal in the thermodynamic limit, we expect that the remaining deviations will disappear if one considers $\mathcal{N} \rightarrow \infty$ and vanishingly weak HW potential $R_{ext}\rightarrow \infty$ or $V_{ext}\rightarrow 0$. 

A further confirmation of our interpretation comes from the overlap between the numerical eigenstates and the Jacks trial wave functions: as one can see in Table \ref{tab:overlaps}, the overlap is very good with the edge Jacks of \eqref{eq:edgeJacks}. This overlap gets even closer to 1 when the Jacks of \eqref{eq:Simon_wf} are considered~\footnote{Note that EJs with EPs of the form \eqref{eq:k_edge_partition} with $k=1$ correspond to single monomials $\mathcal{M}_{\eta}$'s independently on the Jack parameter and therefore for the $k=1$ energy branch wave functions \eqref{eq:Simon_wf} and \eqref{eq:edgeJacks} coincide.} and/or when a weaker confinement is considered.

\begin{table}
\begin{tabular}{c | c | c | c}
EP      &   $\braket{J^{\beta}_{\eta} \psi_{L}}{\Psi}$			&	$\braket{J^{\nu}_{\eta} \psi_{L}}{\Psi}$	&	$\braket{J^{\beta}_{\eta} \psi_{L}}{J^{\nu}_{\eta} \psi_{L}}$ \\
\hline
$\eta = [1,1,1,1]$		 	&	0.9991						&	0.9991						&	1.0000	\\
						&      	\textcolor{red}{0.9994}			&	\textcolor{red}{0.9994}			&			\\
\hline
$\eta = [2,1,1]$		 	 	&	0.9783						&	0.9936						&	0.9878	\\
						&      	\textcolor{red}{0.9798}			&	\textcolor{red}{0.9953}			&			\\
\hline
$\eta = [2,2]$				&	0.9896						&	0.9902						&	0.9908	\\
						&	\textcolor{red}{0.9914}			&	\textcolor{red}{0.9922}			&			\\
\hline
$\eta = [4,4]$				&	0.9473						&	0.9501						&	0.9452	\\
						&      	\textcolor{red}{0.9669}			&	\textcolor{red}{0.9733}			&			\\
\end{tabular}
\caption{Overlaps of the numerical ED eigenstates $\ket{\Psi}$ with the corresponding trial wave functions \eqref{eq:edgeJacks} and \eqref{eq:Simon_wf} for a system of $\mathcal{N}=6$ particles. Values in black refer to a HW confinement of parameters $R_{ext} = 5.25 \sqrt{2} l_{B}$ and $V_{ext} = 30 \, V_{0}$ while those in red to the $R_{ext} = 5.25 \sqrt{2} l_{B}$ and $V_{ext} = 15 \, V_{0}$ case.}
\label{tab:overlaps}
\end{table}

%------------------------------------------------------------------------------------------------
\subsection{\label{subsec:sub-branches}Order of sub-branches}

\begin{figure*}[t!]
\includegraphics[width=1\textwidth]{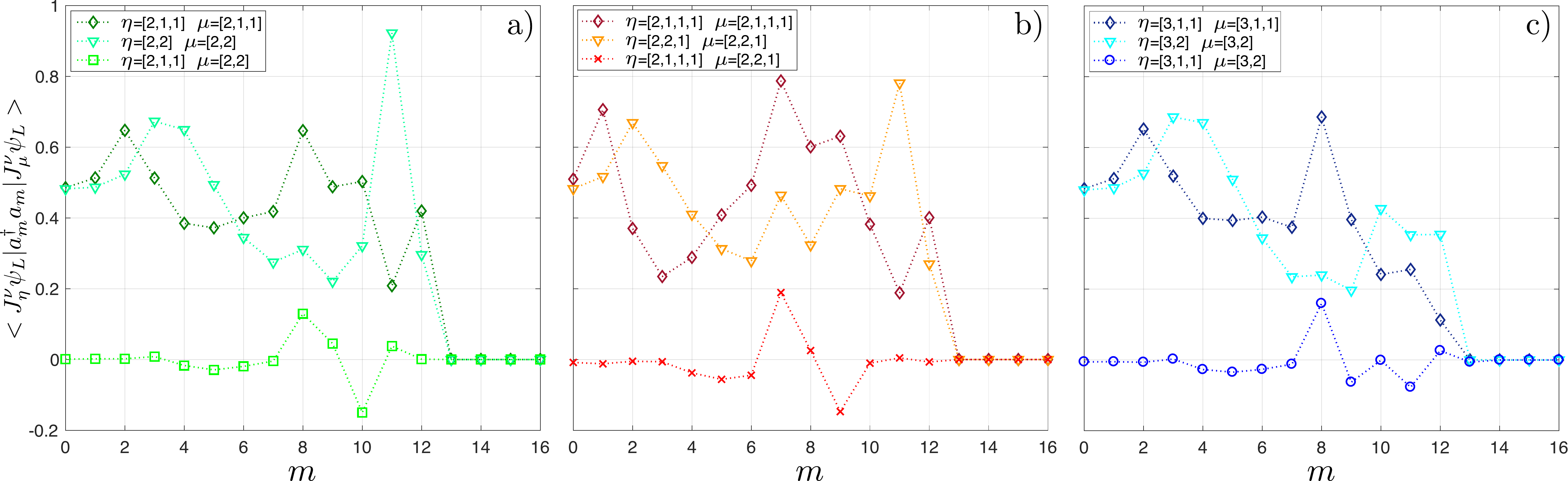}
\caption{Matrix elements of the operators $a^{\dagger}_{m} a_{m}$ evaluated on trial wave functions \eqref{eq:Simon_wf} describing states having the same angular momentum eigenvalue and belonging to the same energy branch but different sub-branches. In particular: panel a) shows results for EJs with EPs $[2,1,1]$ and $[2,2]$, panel b) concerns EJs with EPs $[2,1,1,1]$ and $[2,2,1]$ and finally panel c) reports $a^{\dagger}_{m} a_{m}$ matrix elements evaluated by considering EJs with EPs $[3,1,1]$ and $[3,2]$. In all cases the off-diagonal matrix elements in the Jack basis are much smaller than the diagonal ones and also characterized by an oscillating dependence on $m$. Both these features lead to very small values once averaged on the different $m$'s and provide an explanation for the unexpected energy crossing with no mixing that take place when passing from our realistic potential regime to the extremely steep limit of~\cite{FernSimon}.}
\label{fig:matrixelements}
\end{figure*}

In the previous sub-section we have presented a general criterion for the ordering in energy of the different branches and of the different sub-branches within a given branch. As clearly shown in Fig. \ref{fig:soft_dens} a) sub-branches associated with higher values of the second EP element $\eta_{2}$ have increasing energies. For instance among the two $L = 34$ eigenstates in the second energy branch [see Fig. \ref{fig:soft_dens} b)], the one in the $\eta_{2} = 1$ sub-branch has a lower energy than the one in the $\eta_{2} = 2$ sub-branch. 

Despite this behavior persists for all HW parameters values we have considered, it is crucial to note that this result is in contrast to what was found in~\cite{FernSimon} for the extremely steep HW limit where sub-branches associated with increasing $\eta_{2}$ values are characterized by decreasing energies.

This disagreement is easily understood by noting how the energy shifts in the extremely steep HW limit only depend on the highest $m$ occupation number, while, as we have seen, this is no longer the case for more realistic confining potentials: in particular, the observed ordering of the sub-branches in the two cases is  explained by the fact that increasing values of $\eta_{2}$ correspond to slightly lower occupations of the highest $m$ orbital but also much higher occupations of the second highest one [see Fig. \ref{fig:occ_num} h)].

This energetic behavior characterizing sub-branches belonging to the same energy branch, together with the observation that the eigenstates are the same in different confining regimes, suggests that in the transition between the two regimes the eigenstates in the different sub-branches cross in energy without mixing. While such a crossing would be obviously protected by rotational symmetry for eigenstates with different angular momenta, it is quite unexpected for same $L_{z}$ eigenstates which would typically mix in a non-trivial and potential-dependent way. 

From our calculations, it however appears that this is not the case and the Jacks \eqref{eq:Simon_wf} remain precise eigenstates of the Hamiltonian for all considered confinement potentials. To further corroborate this statement,  we have studied the value of the matrix elements of the ${a}^\dagger_m {a}_m$ operators contributing to the confinement energy \eqref{eq:confinement2}: as one can see in Fig. \ref{fig:matrixelements}, their off-diagonal matrix elements in the Jack basis turn out to be much smaller in magnitude than the diagonal ones and, even more remarkably, have a markedly oscillating dependence on $m$. 

As a result, they easily average to very small values when realistic potentials $\mathcal{U}_m$ are considered. For instance, for the realistic confinement potential considered in Fig.\ref{fig:soft_dens}(a), the rescaled expectation value,
\begin{equation}
\frac{\langle \phi_\eta | H_{conf} | \phi_\mu\rangle}{\sqrt{[\langle \phi_\eta | H_{conf} | \phi_\eta\rangle\langle \phi_\mu | H_{conf} | \phi_\mu\rangle]}}
\end{equation}
takes respectively the (very small) values $0.0033$, $0.0202$ and $0.0119$ for the $\phi_\eta$ and $\phi_\mu$ wave functions corresponding to the Jacks considered in the three panels of Fig. \ref{fig:matrixelements}.

%------------------------------------------------------------------------------------------------
\subsection{\label{subsec:QH}Fine structure of the spectrum}

In the previous subsections we have proposed a hierarchical classification of the states in terms of the first entries of the EJ root partition. While this criterion allows to classify the order in energy of the different branches and sub-branches, it does not offer much insight on the physical mechanisms underlying the relative energy of states within a given branch or sub-branch. As the derivative of the energy dispersion with respect to angular momentum determines the angular speed of a perturbation propagating along the edge of the disk-shaped cloud, an experimental measurement of the surface dynamics and its collective modes may provide a valuable insight on the nature of the edge excitations and on its deviation from standard chiral Luttinger liquid theory~\cite{Xiao-GangWen}.

A complete physical interpretation of the numerical results is a very complicate task, in this section we mainly focus our attention on the possibility of having the single-QH excitation state
\begin{equation}
\psi_{1-qh} (\{ z_{i}\}, \xi = 0 ) \propto \bigg( \prod_{i} z_{i} \bigg) \psi_{L}(\{ z_{i}\}) .
\end{equation} 
as the first excited state above the Laughlin ground state. The interest of this specific feature stems from its marked deviation from the chiral Luttinger liquid theory of edge states (which predicts a monotonical increase of the eigenenergies as a function of $\Delta L$) as well as from the potential utility of a massive population of QH states in view of studies of anyon physics. In Fig.\ref{fig:soft_spectra} d), we illustrate a set of confinement parameters for which this is indeed the case.

This peculiar energetic behavior of the lowest energy excited states in the weak confinement regime can be explained by looking at the expectation value of occupation numbers $\bar{n}_{m} \equiv \expval{a^{\dagger}_{m} a_{m}}{\psi}$ taken on trial wave functions corresponding to states in the $k=1$ energy branch. While for stronger confinement potentials one should take into account the reorganization of the states within the manifold of non-interacting states as well as the possible mixing with quasi-particle states above the Laughlin gap, such a simple analysis based on the occupation numbers is expected to be accurate at a linear response level in the weak confinement limit.

\begin{figure}[ht!]
\includegraphics[width=0.504\textwidth]{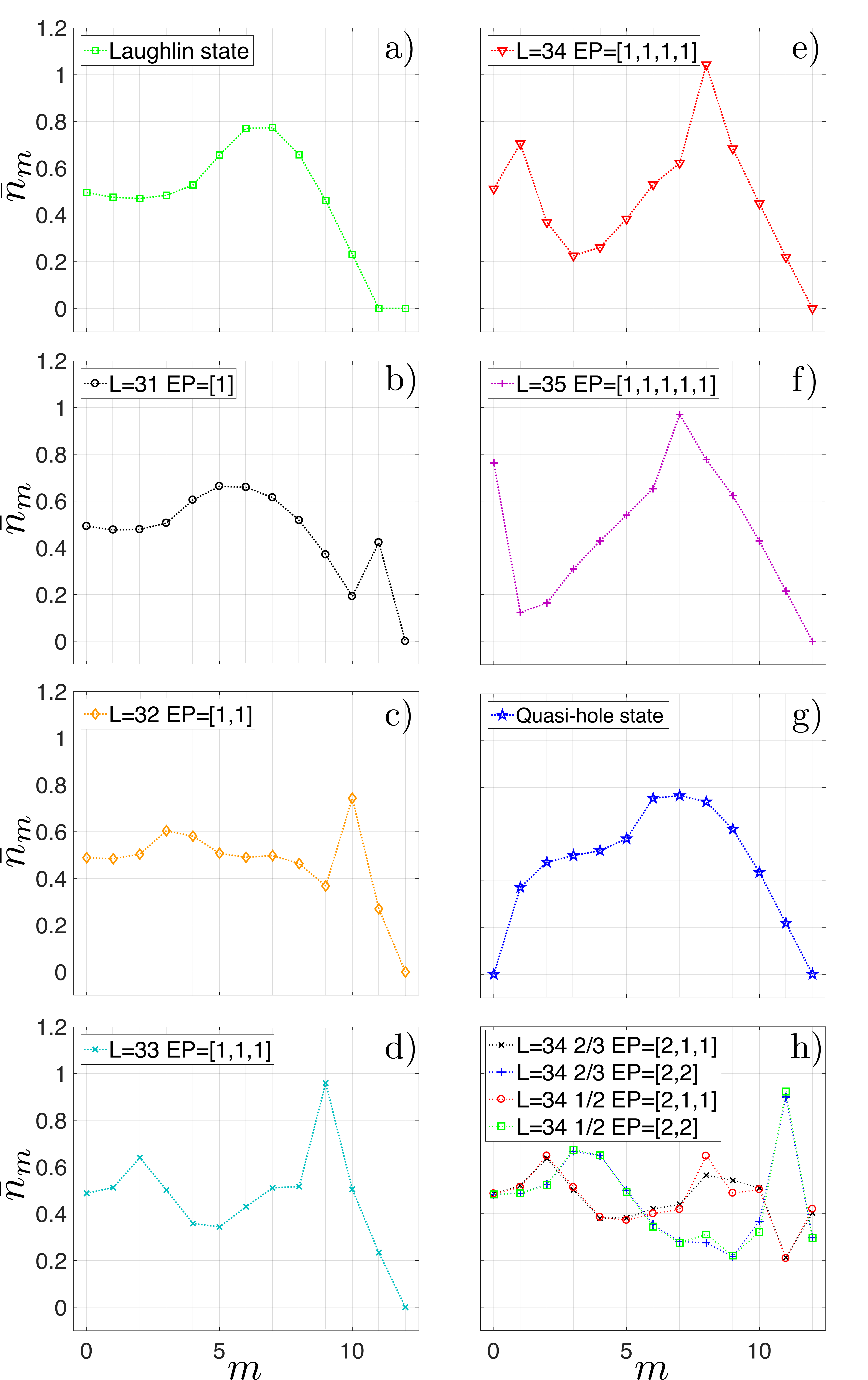}
\caption{Panels from a) to g) show the average occupation numbers $\bar{n}_{m} \equiv \expval{a^{\dagger}_{m} a^{}_{m}}{\psi}$ of single-particle LLL orbitals calculated on Jack trial wave functions for states in the $k=1$ energy branch [38]. The markedly different shapes of $\bar{n}_{m}$ shown in panels a)-g) for the different states  within the lowest $k=1$ energy branch can be used to explain their ordering in energy. In panel h) we compare the $\bar{n}_{m}$ of states within the $k=2$ energy branch as predicted by Jack trial wave functions with different $\beta=2/3$ and $\nu=1/2$ parameters: the differences are minor and the global behavior of $\bar{n}_{m}$ as function of $m$ turns out to be mostly determined by the EP. In particular, trial wave functions with EP's $[2,2]$ and $[2,1,1]$ are characterized by very different values of $\bar{n}_{m}$ for $m=11,12$. Together with the fact that the potential energy is dominated by the highest-$m$ $\mathcal{U}_{m}$, this explains the different ordering of sub-branches found in our calculations compared to the extremely steep limit of~\cite{FernSimon}.}
\label{fig:occ_num}
\end{figure}

The distribution of the occupation numbers $\bar{n}_m$ among different $m$ single particle orbitals for growing total angular momentum in the first $k=1$ branch of excited states is shown in Fig. \ref{fig:occ_num}. As we increase $\Delta L$, the broad central peak of occupation visible around $m\simeq \mathcal{N}=6$ for the $\Delta L=0$ Laughlin state slowly moves towards lower $m$'s while becoming less pronounced. At the same time, another peak appears at the high-$m$ edge of the distribution and moves towards lower $m$ while becoming more pronounced until it transforms into a central peak in the single-QH state. Remarkably, the $m$-distribution of this latter state is quite similar to the one of the Laughlin state, just shifted by one unit of $m$. Even though the features of the $m$-distribution shift towards small $m$'s for growing $\Delta L$, the overall center of mass of the distribution grows as expected as $\Delta L$.

From these curves, we can notice that values of the highest $m$'s occupation numbers - i.e. $\bar{n}_{9}$, $\bar{n}_{10}$ and $\bar{n}_{11}$ - are very similar for the single-QH state and for the neighboring Jack with EP $\eta = [1,1,1,1,1]$ and $L=35$. Actually those occupation numbers are slightly lower for this latter state. This explains why for large enough $R_{ext}$ the energy shift due to the HW potential is almost the same for these two states and in particular why the $ L = 35$ edge state is -by a short difference- the first excited state. Indeed for very large values of $R_{ext}$ all contributions to \eqref{eq:confinement2} from lower $m$ states can be neglected and the occupation numbers of the highest $m$ state fully determine the confining energy, as discussed in detail in~\cite{FernSimon}.

On the other hand, it is also possible to obtain spectra in which the single-QH state is the first excited one, as shown in Fig. \ref{fig:soft_spectra} d). This happens when -at fixed $V_{ext}$- we slightly reduce the HW radius $R_{ext}$ and can be attributed to the fact that the $\eta = [1,1,1,1,1]$ Jack polynomial has higher $\bar{n}_{m=8,9}$ values than the single-QH state.  As a result, for lower values of $R_{ext}$ the energy contribution given by these $m$'s may overcome the one relative to the higher $m$'s, so that the energy shift caused by the HW confinement on the $L=35$ edge state may be bigger than that on the single-QH state. 
Fig. \ref{fig:occ_num} shows also that for $m<7$ the single-QH state has values of the occupation numbers which are higher than those of the $\eta = [1,1,1,1,1]$ Jack polynomial. Therefore, in light of what we have just said, we expect that for even smaller values of $R_{ext}$ the single-QH state will cease to be the first excited one in favor of lower angular momentum edge states.
\begin{figure}[t]
\includegraphics[width=0.48\textwidth]{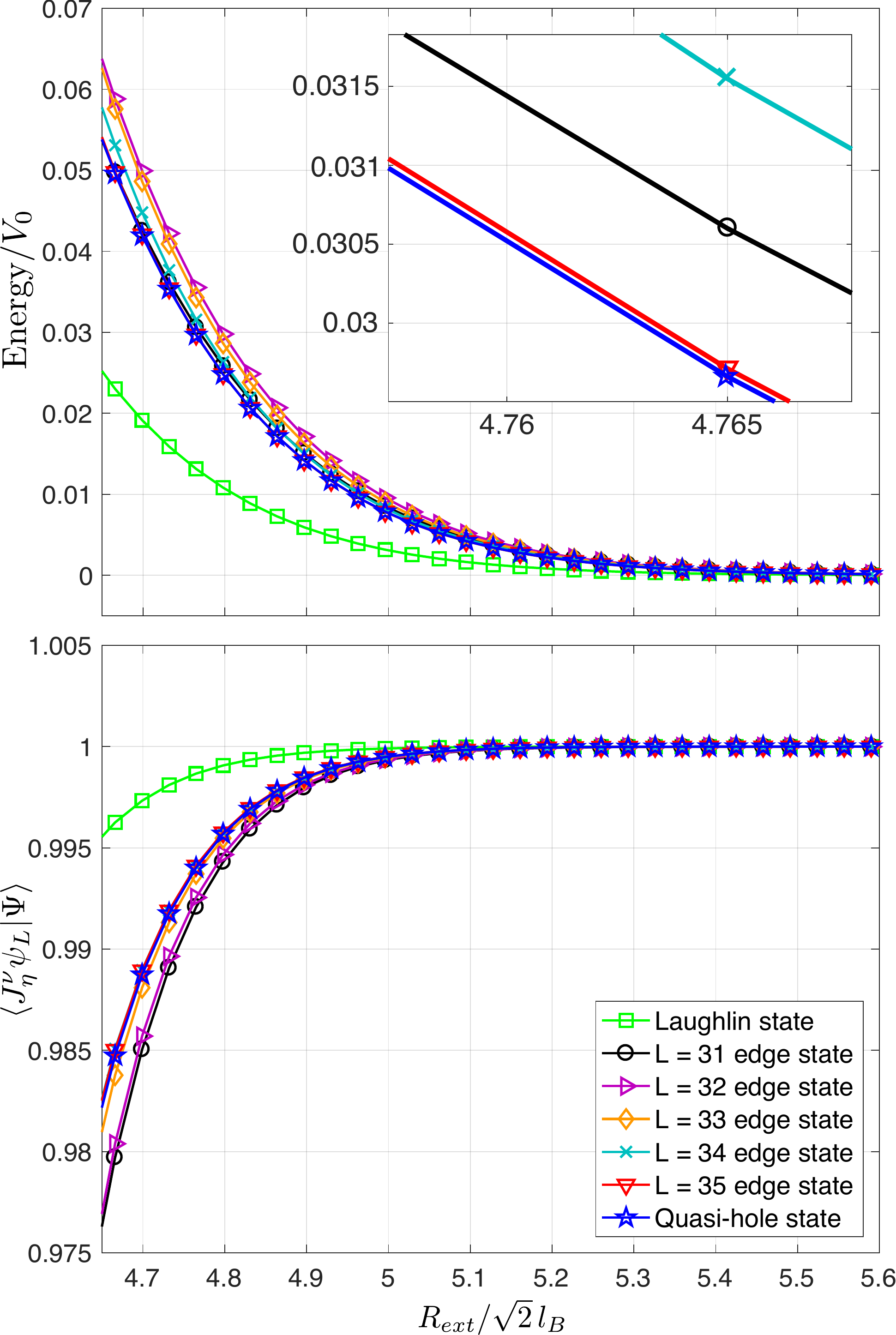}
\caption{Upper panel: energetic behavior of the Laughlin state and its single-QH and edge excitations as function of the HW radius $R_{ext}$ for fixed values of the HW strength $V_{ext} = 10 \, V_{0}$ and of the particle number $\mathcal{N} = 6$. The inset shows the existence of an interval in which the single-QH state is the first excited state. Lower panel: overlap between the exact numerical eigenstates and the Jack trial wave functions describing the 1st energy branch. Despite for large values of $R_{ext}$ the overlap between the lowest energy L=36 eigenstate and the single-QH wave function is of the order of $1$, it remains very high also for $R_{ext}$ values for which the L=36 state is the first excited one.}
\label{fig:soft-EoverR}
\end{figure}

This physics is summarized in the plots of the $R_{ext}$-dependence of the energies of the Laughlin state and of its QH and edge excitations that are shown in the upper panel of Fig.\ref{fig:soft-EoverR}~: for a fixed value of $V_{ext}$ there exists in fact a finite interval of $R_{ext}$ values in which the single-QH state is indeed the first excited one. At the same time, it is important to stress the fact that despite for these trap parameters the EE energies are not negligible respect to the Laughlin gap ($0.1 \, V_{0}$), the overlap between numerical eigenstates and $\eta = [1, \dots]$ Jack polynomials remain extremely high [see Fig.\ref{fig:soft-EoverR}, lower panel] meaning that the first excited state is really the single-QH one.

Also the energetic behavior characterizing the other edge states - i.e. those with $ 32< L < 35$ - can be explained in terms of the occupation numbers plotted in Fig. \ref{fig:occ_num}. Indeed we can note that Jacks describing states in the $k=1$ energy branch associated with increasing values of the additional angular momentum $\Delta L$, are characterized by occupation numbers which show peaks at lower $m$'s. As for sufficiently large $R_{ext}$, the dominant contribution to $H_{conf}$ comes from large $m$ terms corresponding to larger $\mathcal{U}_{m}$ values, it is completely reasonable that higher $\Delta L$ states have lower energies, in agreement with~\cite{FernSimon}. Quite strikingly, note that a decreasing energy with $\Delta L$ means that wavepackets of edge excitations propagate backwards with respect to the cyclotron orbits, in disagreement with the usual chiral Luttinger liquid theory.

Despite this explanation is in complete agreement with what we observe for $ L \geq 32$, it seems to be odd with the energetic behavior of the $ L = 31$ Hamiltonian eigenstate corresponding to the global dipole-like motion of the cloud. The Jack associated with such a states has EP $\eta = [1]$ and it has the highest occupation of the $m=11$ LLL wave function. However, this Jack has the peculiarity of having all the other high $m$ occupation numbers lower than those of both the single-QH state and the $\eta = [1,1,1,1,1]$ Jack, which explains the observation of energies for the $L=31$ state which are similar to the ones of the $L=35$ and of the single-QH states.

%{\color{red}As one can see in the left part of Fig.\ref{fig:soft-EoverR}, the situation is completely different for very low values of $R_{ext}/\sqrt{2} l_B\lesssim 4$ when $\mathcal{U}_{m}$ are a weakly growing function of $m$. In this regime, the energy of the states within the $k=1$ branch monotonically increases with the angular momentum $L$. This recovers a qualitative behaviour analogous to the chiral Luttiger liquid theory, with surface excitations propagating along the direction of the cyclotron orbits. Still, a significant splitting to the other branches persists.}

Starting from these intriguing numerical results, on-going work is trying to understand the physical origin of the 
deviations from the chiral Luttinger liquid theory, so to disentangle finite-size effects and highlight the interesting edge physics, including nonlinear effects in the edge dynamics that were anticipated in~\cite{Wiegmann} and may be responsible for the mixing of different quantum states of the Luttinger liquid.

%------------------------------------------------------------------------------------------------
\subsection{\label{subsec:compressibility}(In)compressibility of the states}

As a further interesting feature of the Laughlin state and of its low-lying excited states, it is interesting to complete our study with a short discussion of their response to an increase of the HW potential strength $V_{ext}$ as a way to measure their compressibility. In view of future experimental studies particular, this strategy may provide access to one of the most celebrated properties of FQH liquids.

\begin{figure}[b]
\includegraphics[width=0.52\textwidth]{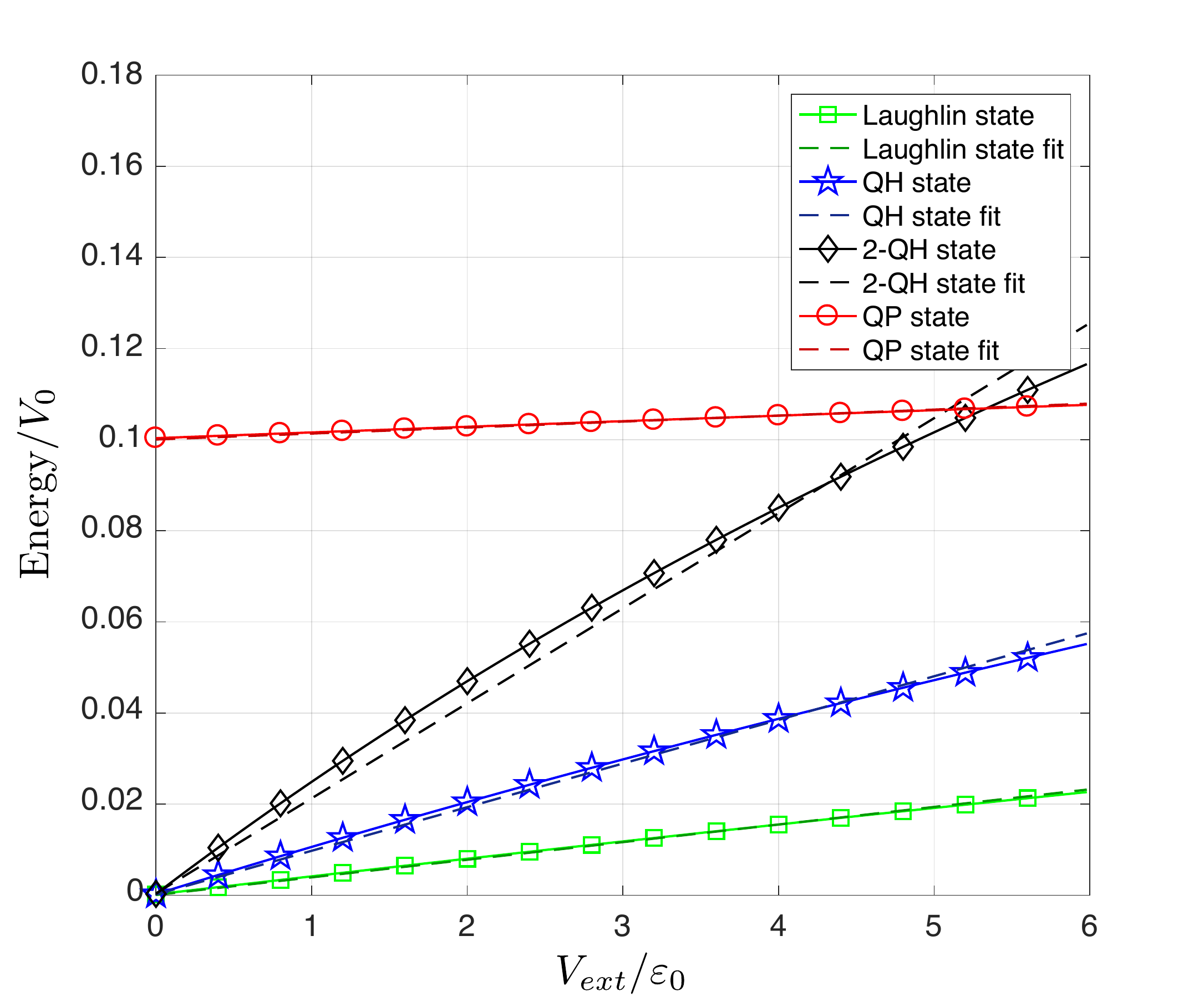}
\caption{Solid lines: energies of the Laughlin state and its QH and QP excitations as function of the HW potential strength $V_{ext}$ for fixed values of the potential radius $R_{ext} = 4.65 \sqrt{2} \, l_{B}$ and of the number of particles $\mathcal{N} = 5$. Dashed lines are linear fits: the more accurate this fit, the weaker the compressibility of the state. As expected, the compressibility dramatically increases when QH's are inserted in the fluid, while it is not affected by the initial presence of QP's.}
\label{fig:soft-EoverV}
\end{figure}

As one can observe in Fig. \ref{fig:soft-EoverV}, the energies of the Laughlin state and of the quasi-particle (QP) state grows almost linearly in the external potential strength, confirming the expected incompressible behavior.

\begin{figure*}[t]
\includegraphics[width=1.0\textwidth]{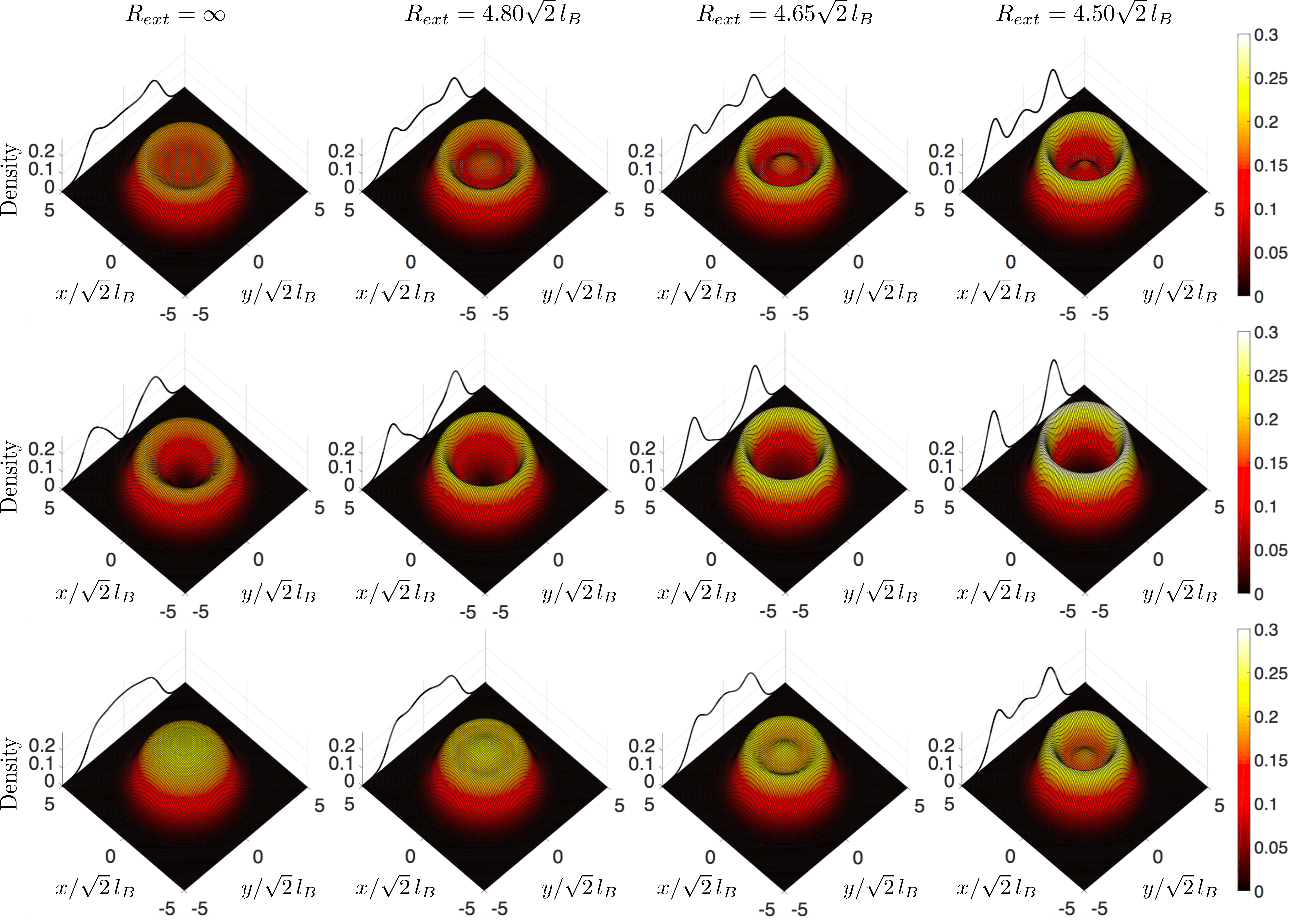}
\caption{Behavior of the Laughlin state (first row), the single-QH state (second row) and the single-QP state (third row) density profiles as function of the HW radius $R_{ext}$ for fixed potential strength $V_{ext} = 100 \, V_{0}$ and particle number $\mathcal{N} = 6$. As we can see, more we reduce the potential radius $R_{ext}$ - from left to right -, more the densities of the Laughlin state and its excitations decrease at the center. Densities are normalized such that their spatial integrals over the whole 2D plane recovers the total number of particles $\mathcal{N}$.}
\label{fig:strong-densities}
\end{figure*} 

On the other hand, the energies of QH excited states grows less than linearly as a function of $V_{ext}$, which witness the ability of these states to rearrange themselves in response to the confinement. This is manifestation of their finite compressibility of the state. As expected, the larger the number of QHs, the stronger this compressibility. A similar compressible behavior is also found for EE states (not shown in the figure).

While these results are restricted to relatively weak confinement potentials that are not able to generate a massive number of extra quasi-particles in the fluid, in the next Section we will see how a strong compression of the cloud is able to distort the density profile of the cloud in quite unexpected ways.

%------------------------------------------------------------------------------------------------
%------------------------------------------------------------------------------------------------
\section{\label{sec:strong}Strong confinement regime}

After having discussed the weak confinement regime where the physics takes place within the non-interacting state manifold, we now turn our attention to the \emph{strong confinement regime} where significant mixing with quasi-particle states above the Laughlin gap can occur and the density distribution of the FQH droplet is sizably compressed in space.

With no loss of generality, we focus on HW potential strengths $V_{ext}$ of the same order as before, but much smaller disk radii $R_{ext}$. For such high values of $V_{ext}$, the strong confinement condition can be reached as soon as $R_{ext} \gtrsim R_{cl}$. The study of this case exhausts the range of confinement regimes and completes the physics of a FQH liquid confined in a HW potential.

For sufficiently strong confinements, the lowest energy many-body states have angular momentum eigenvalues lower than the $\nu = 1/2$ Laughlin state one $L_{L} = \mathcal{N} (\mathcal{N}-1)$: Even though for no (or weak) confinement the energies of all eigenstates of low angular momentum $L < L_{L}$ lie above the bulk Laughlin gap, the confinement potential has in fact a much weaker effect on these states than on the spatially more extended states of the Laughlin family, so their relative order in energy can be swapped.

This physically expectable result is accompanied by the surprising behavior of the density profiles of the eigenstates that is illustrated in Fig. \ref{fig:strong-densities} for the Laughlin state and its QH and QP excited states: the more one squeezes the system by reducing $R_{ext}$, the more the value of the density at the center decreases. This apparently counterintuitive behavior can be explained if one takes into account the invariance under rotation of the Hamiltonian and the associated conservation of the total angular momentum. 

Each eigenstate of the Hamiltonian can in fact be written as a linear combination of elements of the occupation number basis \eqref{eq:linear_comb} sharing the same angular momentum. As a result, a large population in high $m$ orbitals must be compensated by a high population in the low $m$ orbitals as well. Since the effect of the confinement potential is strongest on the high-$m$ orbitals, minimization of the confinement energy leads to a reorganization of the eigenstates in favor of those configurations that show a reduced occupation of high-$m$ orbitals and, consequently, of low-$m$ orbitals as well. As the density at the center of the cloud is mostly determined by low-$m$ states, the mechanical compression of the droplet from outside leads to a marked depletion of the central region as well, as visible in the $r\approx 0$ region of the right-most panels of Fig. \ref{fig:strong-densities}.

%------------------------------------------------------------------------------------------------%------------------------------------------------------------------------------------------------
\section{\label{sec:conclusions}CONCLUSIONS}

In this work we have made use of numerical exact diagonalization calculations to study the effect of a realistic hard-wall confinement on a $\nu = 1/2$ fractional quantum Hall droplet of bosons with contact interactions. 

We have found that the physics of such systems is far richer than that of harmonically confined systems and of HW models in the extremely steep limit considered in~\cite{FernSimon}. The more general and realistic potential we have considered removes the degeneracy between edge excitations with the same angular momentum, so that many-body states organize themselves in energy branches that can be interpreted in terms of Jack polynomials. The energy ordering of states can be controlled via the potential parameters in a much wider way than originally expected on the basis of a chiral Luttinger liquid theory. In particular, parameters for which the single-quasi-hole state is the first excited state above the Laughlin ground state have been identified and level crossings unexpectedly with no mixing have been found. Under a strong confinement potential, the density profile of the droplet results dramatically modified with the surprising appearance of a density depression at the center. Interesting experimental consequences of our theoretical results have also been discussed.

Together with the extremely precise one-to-one correspondence between the energy eigenstates and the Jacks trial wave functions, these features make a weak HW confinement one of the most appealing regimes where to perform experimental investigation of fractional quantum Hall physics in atomic of photonic systems. Ultracold atoms in different traps and photons in suitably designed cavities or cavity arrays are in fact the subject of intense studies as quantum simulators of many-body physics as well as prospective candidates where to observe and study advanced topological many-body states, potentially with quantum information applications.

From a theoretical point of view, the description of eigenstates in terms of Jack polynomials appears to be a powerful tool in view of larger scale numerical calculations, but may also contribute to the physical understanding of unexpected features such as the the energy anticrossing discussed in \ref{subsec:sub-branches}. Furthermore, it could help optimization of protocols to generate fractional quantum Hall states in a driven-dissipative context as recently proposed in~\cite{OUIC_IncoherentFQH}, to understand the physical origin of the marked deviation of the numerically observed spectra from the chiral Luttinger liuqid theory, to unravel the response of FQH droplets to time-dependent potentials, as well as to investigate more complex ring geometries in which excitations of the inner and outer edges can be strongly mixed and quasi-hole tunneling phenomena can occur. All these issues are the subject of on-going research.

%------------------------------------------------------------------------------------------------%------------------------------------------------------------------------------------------------
\section{\label{sec:acknowledgments}ACKNOWLEDGMENTS}

Continuous discussions and early stage technical support from R. O. Umucal\i lar are warmly acknowledged. This work was supported by the EU-FET Proactive grant AQuS, Project No. 640800, and by the Autonomous Province of Trento, partially through the project ``On silicon chip quantum optics for quantum computing and secure communications" (``SiQuro").  IC acknowledges the Kavli Institute for Theoretical Physics, University of California, Santa Barbara (USA) for the hospitality and support during the early stage of this work.

\bibliography{bibliography}
   
\end{document}